\documentclass[journal=jacsat,manuscript=article]{achemso}

\usepackage[version=3]{mhchem} % Formula subscripts using \ce{}
\usepackage{float}
\usepackage{caption}
\usepackage{subcaption}
\usepackage{hyperref}
\pdfoutput=1

\author{Tunan Xia}
\affiliation[Pennsylvania State University]
{Department of Electrical Engineering, Pennsylvania State University, 210 Old Main, University Park, PA 16802, USA}

\author{Cheng-Kuan Wu}
\affiliation[National Sun Yat-Sen University]
{Department of Photonics, National Sun Yat-Sen University, No. 70 Lien-hai Road, Kaohsiung 80424, Taiwan ROC}

\author{Duan-Yi Guo}
\affiliation[National Sun Yat-Sen University]
{Department of Photonics, National Sun Yat-Sen University, No. 70 Lien-hai Road, Kaohsiung 80424, Taiwan ROC}
\author{Lidan Zhang}
\affiliation[Pennsylvania State University]
{Department of Electrical Engineering, Pennsylvania State University, 210 Old Main, University Park, PA 16802, USA}
\author{Bofeng Liu}
\affiliation[Pennsylvania State University]
{Department of Electrical Engineering, Pennsylvania State University, 210 Old Main, University Park, PA 16802, USA}
\author{Tsung-Hsien Lin}
\affiliation[National Sun Yat-Sen University]
{Department of Photonics, National Sun Yat-Sen University, No. 70 Lien-hai Road, Kaohsiung 80424, Taiwan ROC}

\author{Xingjie Ni}
\affiliation[Pennsylvania State University]
{Department of Electrical Engineering, Pennsylvania State University, 210 Old Main, University Park, PA 16802, USA}
\author{Iam-Choon Khoo}
\affiliation[Pennsylvania State University]
{Department of Electrical Engineering, Pennsylvania State University, 210 Old Main, University Park, PA 16802, USA}
\author{Zhiwen Liu}
\affiliation[Pennsylvania State University]
{Department of Electrical Engineering, Pennsylvania State University, 210 Old Main, University Park, PA 16802, USA}
\email{zzl1@psu.edu}

\title[An \textsf{achemso} demo]
  {Ensemble nonlinear optical learner by electrically tunable linear scattering}

\abbreviations{IR,NMR,UV}

\begin{document}

% \begin{tocentry}

% Some journals require a graphical entry for the Table of Contents.
% This should be laid out ``print ready'' so that the sizing of the

% Inside the \texttt{tocentry} environment, the font used is Helvetica
% 8\,pt, as required by \emph{Journal of the American Chemical

% The surrounding frame is 9\,cm by 3.5\,cm, which is the maximum
% permitted for  \emph{Journal of the American Chemical Society}
% graphical table of content entries. The box will not resize if the
% content is too big: instead it will overflow the edge of the box.

% This box and the associated title will always be printed on a
% separate page at the end of the document.

% \end{tocentry}

\begin{abstract}
  Recent progress in effective nonlinearity, achieved by exploiting multiple scatterings within the linear optical regime, has been demonstrated to be a promising approach to enable nonlinear optical processing without relying on actual material nonlinearity. Here we introduce an ensemble nonlinear optical learner, via electrically tunable linear scattering in a liquid-crystal-polymer composite film under low optical power and low applied electrical voltages. We demonstrate, through several image classification tasks, that by combining inference results from an ensemble of nonlinear optical learners realized at different applied voltages, the ensemble optical learning significantly outperforms the classification performance of individual processors. With very low-level optical power and electrical voltage requirements, and ease in reconfiguration simply by varying applied voltages, the ensemble nonlinear optical learning offers a cost-effective and flexible way to improve computing performance and enhance inference accuracy.
\end{abstract}

\section{1. Introduction}
Deep neural networks (NNs) have made profound technological impacts in recent years \cite{liSurveyConvolutionalNeural2022, abiodunComprehensiveReviewArtificial2019, dhillonConvolutionalNeuralNetwork2020, schmidhuberDeepLearningNeural2015, wuDevelopmentApplicationArtificial2018, samekExplainingDeepNeural2021, egmont-petersenImageProcessingNeural2002, abiodunStateoftheartArtificialNeural2018, dayhoffNeuralNetworkArchitechtures1990}. However, these advancements also come with great costs in computing resources and energy consumption \cite{thompsonComputationalLimitsDeep2022, lohnAIComputeHow2022, desislavovComputeEnergyConsumption2023, hestnessDeepLearningScaling2017, mcsherryScalabilityWhatCOST2015, rosenfeldScalingLawsDeep2021, kaplanScalingLawsNeural2020}, driven by ever-growing model sizes, increasingly large datasets, and the continuing pursuit of higher performance. Optics and photonics-based computing offers significant advantages in speed, bandwidth, and energy efficiency \cite{millerAttojouleOptoelectronicsLowEnergy2017, caulfieldOpticalNeuralNetworks1989, wetzsteinInferenceArtificialIntelligence2020, kitayamaNovelFrontierPhotonics2019, demarinisPhotonicNeuralNetworks2019, miscuglioPhotonicTensorCores2020, shastriNeuromorphicPhotonicsPrinciples2018}. For example, optical diffractive NNs \cite{linAllopticalMachineLearning2018, yanFourierspaceDiffractiveDeep2019, qianPerformingOpticalLogic2020, zhuSpaceefficientOpticalComputing2022} have demonstrated the capabilities to perform a variety of tasks, including image classification and logical operations with low latency (computing at light speed). Matrix multiplication, arguably the most prevalent operation in NN, can be efficiently implemented optically (e.g., through the use of a mesh of Mach Zehnder interferometers) to accelerate the calculation \cite{pesericoIntegratedPhotonicTensor2023, chengPhotonicMatrixComputing2021, zhouPhotonicMatrixMultiplication2022, wuProgrammableIntegratedPhotonic2024, xuSiliconbasedOptoelectronicsGeneralpurpose2022}. However, a multi-layer linear network is equivalent to a single-layer linear network, which cannot approximate nonlinear functions demanded by deeper architectures. Achieving nonlinearity is essential for realizing a universal approximator and for implementing a deep NN \cite{hornikMultilayerFeedforwardNetworks1989, hornikApproximationCapabilitiesMultilayer1991}.

Optical nonlinearity in materials has been explored in optical NNs. For example, photonic neurosynaptic networks capable of self-learning were demonstrated using phase-changing materials \cite{feldmannAllopticalSpikingNeurosynaptic2019}. A scalable optical learning operator was explored via the nonlinear propagation of picosecond pulses in a multi-mode silica fiber \cite{teginScalableOpticalLearning2021}. Other nonlinear materials, including coupled metallic particles/quantum dot structures \cite{miscuglioAllopticalNonlinearActivation2018}, MXene-Nanoflakes \cite{hazanMXeneNanoflakesEnabledAllOpticalNonlinear2023}, atomic vapor \cite{ryouFreespaceOpticalNeural2021}, and disordered media \cite{wangLargescalePhotonicComputing2024} were also studied. However, the implementation of highly efficient all-optical nonlinearity remains a great challenge. Highly nonlinear materials (e.g., $\chi^{(3)} \geq 10^{-2}$  esu) often suffer from high loss and slow response (e.g., $ms \ -\ s$) \cite{khooNonlinearOpticsActive2014, khooNonlinearOpticsLiquid2009}, while ultrafast nonlinearity typically requires high peak intensities and/or long interaction lengths. These constraints make it difficult to meet the criteria for practical implementations. Optoelectronic approaches have also been investigated to realize effective nonlinearity \cite{psaltisOptoelectronicImplementationsNeural1989, lentineEvolutionSEEDTechnology1993, hamerlyLargeScaleOpticalNeural2019, shenDeepLearningCoherent2017, taitSiliconPhotonicModulator2019, williamsonReprogrammableElectroOpticNonlinear2020, ashtianiOnchipPhotonicDeep2022, pourfardExperimentalRealizationArbitrary2020, wangImageSensingMultilayer2023},  which typically involves the optical-to-electronic conversion and the accompanying detection nonlinearity, followed by the electronic-to-optical conversion. Limitations of these methods include increased latency and energy consumption, lower speed, and additional complexity/cost in scaling the architecture.

Recently, a new paradigm to achieve effective nonlinearity in the linear optics regime \cite{eliezerTunableNonlinearOptical2023, xiaNonlinearOpticalEncoding2024, yildirimNonlinearProcessingLinear2024, liNonlinearEncodingDiffractive2024, wanjuraFullyNonlinearNeuromorphic2024} has been proposed, where digital data is encoded as a scattering potential. Multiple interactions with this information-bearing scattering potential generate a nonlinear mapping between input data and output scattered field $\mathbf{E_{sc}}$, i.e., $\mathbf{E_{sc}} = \mathbf{T}\mathbf{E_{in}}$ where $\mathbf{E_{in}}$ is the input field and $\mathbf{T}$ is the transfer operator as a result of multiple scatterings. The transfer operator is nonlinearly dependent on the structure which mediates the multiple scatterings/interactions, and thus one may call such nonlinearity as structural nonlinearity, as opposed to optical nonlinearity of materials associated with optical field induced second-, third- or higher- order susceptibilities. Note that although the transfer operator is structurally nonlinear, the output is linearly proportional to the input field, and therefore high power/intensity light sources are not needed to perform the nonlinear input-output mapping. This was first demonstrated by Eliezer et al. \cite{eliezerTunableNonlinearOptical2023}, who constructed a reverberating cavity using an integrating sphere to confine light and result in multiple scatterings off a digital mirror device (DMD) that encoded input data, creating a nonlinear mapping between the input data and the output speckle patterns. Xia et al. \cite{xiaNonlinearOpticalEncoding2024} applied the system to various computing tasks such as image classification and key point detection, demonstrating enhanced learning performance and efficient optical information compression. Yildirim et al. \cite{yildirimNonlinearProcessingLinear2024} used a multi-bounce, single-pass cavity formed by a liquid crystal spatial light modulator (SLM) and a mirror where light was modulated multiple times by the input data (repeated on the SLM) as it propagated between these two components. They demonstrated enhanced performance achieved by data repetition in classification tasks and showed that the system shares similar scaling laws with digital deep neural networks.

Here we introduce an ensemble learning approach to further enhance nonlinear optical processing, where we create an ensemble of nonlinear optical learners by electrically reconfiguring linear scatterings in an electro-optic liquid-crystal-polymer composite (LCPC)  \cite{guoReconfigurableScatteringLiquid2025, chen2D3DSwitchable2016, chiangSelectiveVariableOptical2018}. Under low optical power and low voltages, each configuration of the LCPC creates a distinct nonlinear optical processor, paired with a trainable single-layer digital NN to perform inference. We demonstrate, through several image classification tasks, that tuning the applied voltage and combining inference results from an ensemble of these processors surpasses the performance of any single configuration. Our studies showed that the optical ensemble learning boosts accuracy by 10 percentage points (from 47.4\% to 57.3\%) when classifying MNIST handwritten digits with only four speckle grains. With very low-level optical power and electrical voltage requirement, and ease in reconfiguration simply by varying applied voltages, the ensemble nonlinear optical learning offers a cost-effective, flexible way to improve computing performance and enhance inference accuracy.

\section{2. Results}

\subsection{2.1 Experimental system}
A schematic diagram of the ensemble nonlinear optical processing system is shown in Fig. 1a. The input data is encoded on a reflection-type SLM and an LCPC cell (backed with a mirror) is positioned against the SLM. A low-power He-Ne laser beam is scattered multiple times between the SLM and the LCPC before being detected by a camera.

The LCPC is fabricated using a conventional phase-separation technique [See Method section] \cite{linMechanismScatteringBistable2021, justiceInterfaceMorphologyPhase2005, liControllingMorphologicalElectrooptical2020}. As shown in the polarized microscope image in Fig. 1b, the LCPC has a disordered and heterogeneous morphology due to mismatched refractive indices of the LC and the polymer. The morphology can be altered by changing the applied AC voltage which reorients the LC director axis and hence the refractive index mismatch, which in turn modifies the scattering characteristics of the LCPC \cite{guoReconfigurableScatteringLiquid2025, linMechanismScatteringBistable2021} [See Supplementary Materials for more details such as the voltage-haze characteristics and the correlation/evolution of scattering patterns]. To demonstrate control of structural nonlinearity using the LCPC, a DMD provides amplitude encodings of 3×3 binary images (producing a total of $2^9$ or 512 patterns). The nonlinear relationship between the output speckle intensity and the input binary data (encoded by the “on” and “off” micro-mirrors on the DMD) can be represented by the Fourier expansion\cite{eliezerTunableNonlinearOptical2023, odonnellAnalysisBooleanFunctions2014}. The Boolean Fourier expansion coefficients then quantify the strengths of various structural-nonlinear orders, in a manner analogous to the nonlinear susceptibility tensors used for characterizing optical nonlinearity of materials.

\begin{figure}[H]
    \begin{subfigure}{\textwidth}
      \centering
      \includegraphics[width=0.8\textwidth]{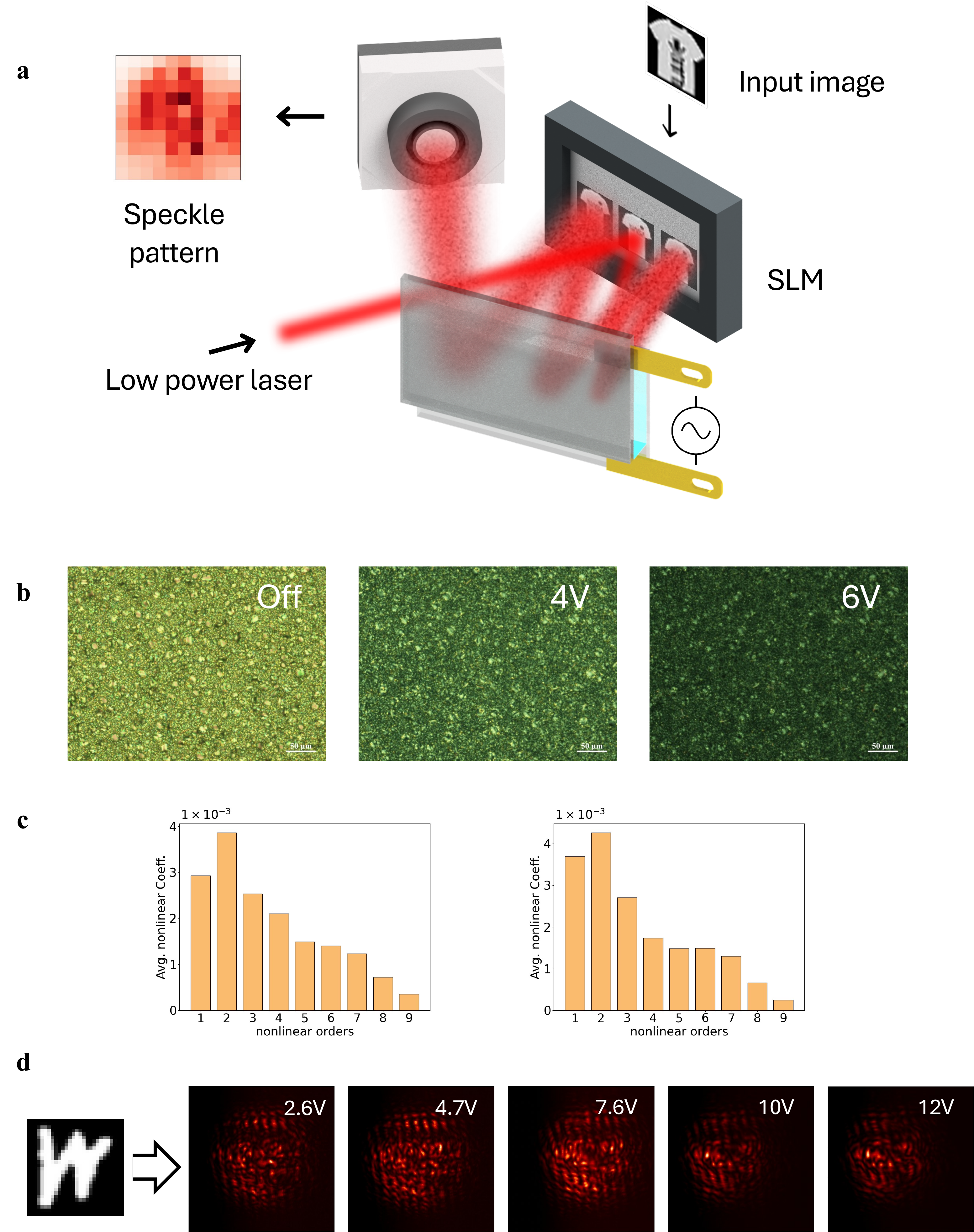}
    \end{subfigure}
  \label{fig:1}
\end{figure}
\clearpage
\begin{figure}[ht]\ContinuedFloat
  \centering
    \caption{\textbf{Ensemble nonlinear optical processing by reconfiguring linear scatterings in liquid crystal/polymer composite (LCPC). (a)} Experimental schematic. A mirror-backed reconfigurable LCPC cell is positioned against a reflection-type SLM that encodes the input data. An alternating current voltage at 1 kHz is applied across the electrodes of the LCPC cell to tune the scattering property. A low-power, continuous-wave He-Ne laser is coupled into the system, and scattered multiple times between the SLM and the LCPC. The output speckle pattern (i.e., feature in a latent space, representing nonlinearly processed input data) is recorded by a camera. \textbf{(b)} Polarized optical microscope images of an LCPC sample under different applied voltages, showing tunable scattering properties. \textbf{(c)} Average structurally nonlinear coefficients of different orders under 0V and 10V applied voltage, respectively; the structural nonlinearity can be tuned by tuning the voltage applied to the LCPC. \textbf{(d)} An example showing a representative input image and its corresponding output speckle patterns under 2.6V, 4.7V, 7.6V, 10V, and 12V applied voltage, respectively; structural nonlinearity based ensemble optical processing can be realized by varying the voltage applied to the LCPC.}
\end{figure}

Fig. 1c compares the average structural-nonlinearity coefficients (i.e., Boolean Fourier expansion coefficients) corresponding to an applied voltage amplitude of 0V and 10V, respectively. With the LCPC operating in the initially ‘opaque’ mode, the system at 0V (stronger scattering) has more weights on higher nonlinear orders while at 10V (weaker scattering) the low-order terms are more pronounced. By adjusting the voltage amplitude applied on the LCPC, a reconfigurable nonlinear mapping between the output light field in the form of a speckle pattern, and the input data is obtained, thus providing a mechanism to enable ensemble learning.

\subsection{2.2 Ensemble nonlinear optical learner}
The ensemble learner is a hybrid optical-digital system. As shown in Fig. 1 (a) the input data is encoded using a phase-only liquid crystal SLM, and the grayscale values (0 - 255) are linearly and proportionally mapped to phase values between 0 and $\pi$. Data encodings are repeated on the SLM at three locations that are aligned with the centers of the scattered beams so that as the laser beam bounces between the SLM and the LCPC cell it interacts with the same input data multiple times to achieve structural nonlinearity. The captured output speckle pattern is then fed into a trainable digital neural network with a single linear layer to yield inference results. Essentially, the optical system processes an input image via multiple linear scatterings and nonlinearly maps the input data into a speckle pattern. It is akin to an extreme learning machine (ELM) with “random” parameters\cite{huangExtremeLearningMachine2006} that are determined by the underlying optical scattering and propagation processes. Note that the parameters of the optical ELM are tunable by adjusting the voltage applied to the LCPC. This reconfigurability creates different optical ELMs for ensemble learning.

The system is evaluated by several image classification benchmarks. In these evaluations, the original grayscale images were encoded into phase images as aforementioned and were up-sampled so that each pixel was mapped to 10×10 SLM pixels, to match the image size with the incident laser beam size. Five representative output speckle patterns are shown in Fig. 1d, which corresponds to the same original image, but different voltage amplitudes applied to the LCPC (2.6V, 4.7V, 7.6V, 10V, 12V, respectively), showing clear differences. The applied electric field is capable of modulating the LCPC scattering properties, which in turn reconfigure the structural nonlinear response of the entire system. In other words, the optical system processes an input image via reconfigurable structural nonlinearity, mapping the input into multiple nonlinearly encoded speckle patterns under different voltages.

The captured speckle patterns were first down-sampled to reduce data size. The down-sampling was guided by the speckle grain size, obtained from the autocorrelation peak widths of the speckle patterns, and determined to be around 32 camera pixels. A super pixel size of 32×32 camera pixels was used. Only a limited number of grains were selected, and this number was controlled by a cropping window centered at the speckle pattern. Grains within the window formed a feature vector, representing the output of the optical system. The dimension of the feature vector varied from 784, which equals the dimension of original images (28×28) from datasets, to only 4 grains. A single-layer linear network that directly connects the input neurons to each class was then trained with the ADAM optimizer on these feature vectors.
\clearpage
\begin{figure}[H]
    \begin{subfigure}{\textwidth}
        \centering
        \includegraphics[width=\textwidth]{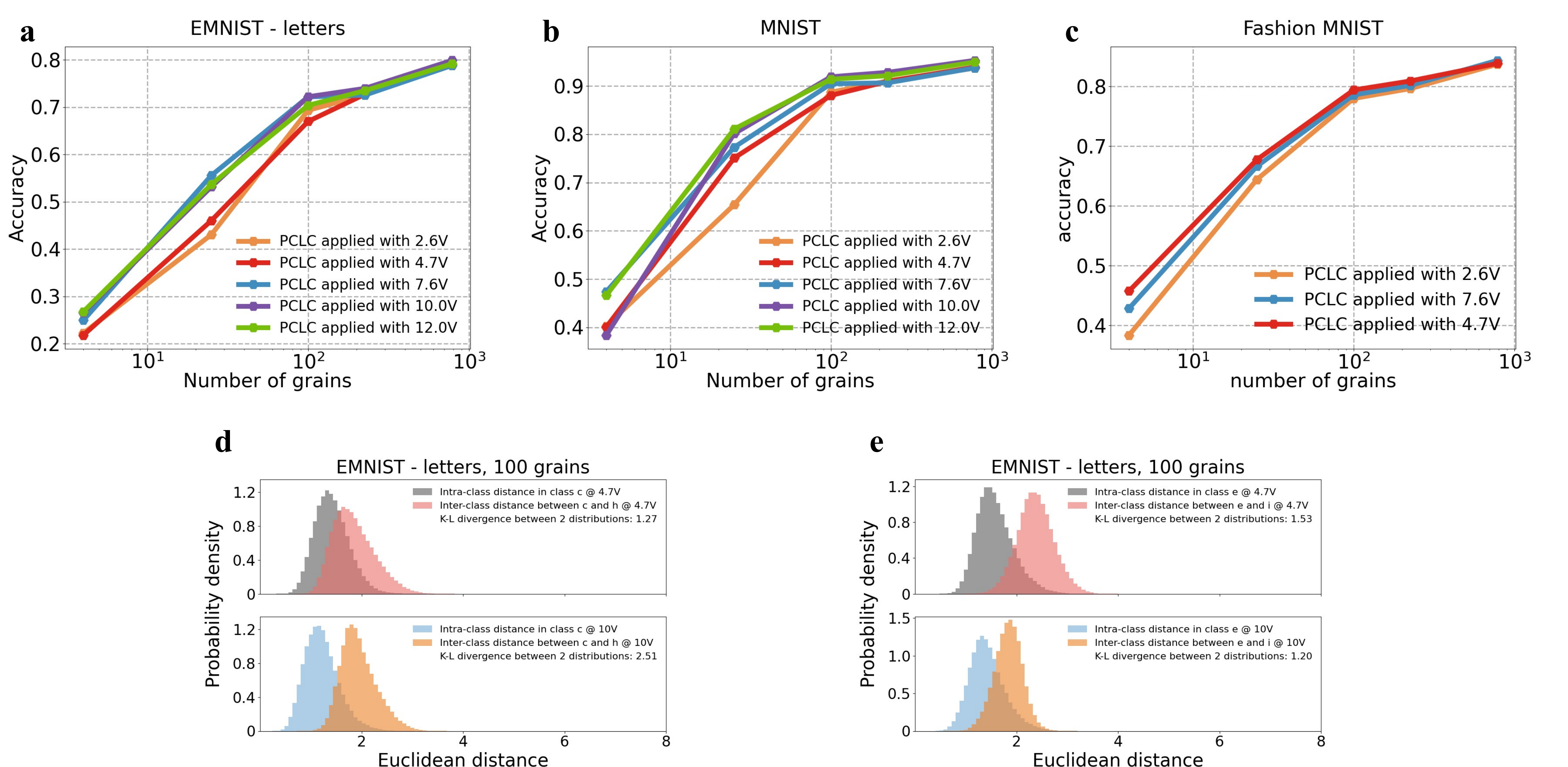}
    \end{subfigure}
    \caption{\textbf{Learning performance in different classification tasks. (a)} EMNIST letters classification accuracies as a function of the number of grains with 2.6V, 4.7V, 7.6V, 10V, and 12V applied to the LCPC, respectively. \textbf{(b)} Accuracy curves for MNIST hand-written digits classification under voltage amplitudes of 2.6V, 4.7V, 7.6V, 10V, and 12V. \textbf{(c)} Accuracy curves for Fashion MNIST classification under 2.6V, 4.7V, and 7.6V. \textbf{(d)} Histograms showing the distributions of intra-class Euclidean distances between feature vectors (with 100 grains) within EMNIST letters dataset class ‘c’ and inter-class Euclidean distances between feature vectors in class ‘c’ and class ‘h’, under 4.7V and 10V. \textbf{(e)} Histograms showing the distributions of intra-class Euclidean distances between feature vectors within EMNIST letters dataset class ‘e’ and inter-class Euclidean distances between feature vectors in class ‘e’ and class ‘i’, under 4.7V and 10V. The distinction between intra- and inter-class distances distributions under the same voltage is quantified by Kullback-Leibler divergence.}
    \label{fig:2}
\end{figure}

Fig. 2 (a-c) also reveal that the classification accuracy depends on applied voltage, particularly at lower dimensions. For instance, in the EMNIST test (Fig.2a) with 4 grains the applied voltage amplitude of 12V yields the best performance ($26.8\%$), but when the number of grains is higher ($\ge25$), 7.6V and 10V perform better, achieving $55.6\%$ (7.6V) with 25 grains, and reaching $72.2\%$ (7.6V and 10V) with 100 grains. At larger grain numbers (beyond 100), accuracies across different voltages converge. Similar behavior can also be seen in the MNIST and Fashion MNIST tests, where certain voltages outperform other voltages at a given grain number. Some insight into these performance variations could be gained through analysis of the Euclidean distances between feature vectors (Fig. 2d, Fig. 2e, see also supplementary materials). Fig. 2d shows the distinction between intra-class (between speckle patterns within class ‘c’) and inter-class distance distributions (between speckle patterns in class ‘c’ and class ‘h’) is greater at 10V (K-L divergence 2.51) than at 4.7V (K-L divergence 1.27). On the other hand, Fig. 2e reveals an opposite trend between class ‘e’ and class ‘i’, where 4.7V (K-L divergence 1.53) slightly outperforms 10V (K-L divergence 1.20) in terms of class separation. These results suggest that the applied voltage can tune the intra- and inter-class distance distributions and therefore tune the classification performance. This unique property not only provides an optimization mechanism to enhance the inference accuracy for a given task, but also enables ensemble learning by taking full advantage of the diversity of the nonlinear processing at all applied voltages.

A conceptual schematic for optical ensemble learning using the reconfigurable structural nonlinearity is illustrated in Fig. 3a: a series of voltages $V_1$,$V_2$,…,$V_n$ is applied to the LCPC to alter the structural nonlinearity by reconfiguring linear scattering in the LCPC; at each applied voltage a corresponding single-layer readout network is trained to classify the speckle patterns produced. When an unseen input is presented, n prediction results are thus obtained (one per applied voltage). Similar to the ensemble techniques in digital machine learning, which combine predictions from multiple base models to enhance the overall performance beyond any single constituent model can achieve\cite{opitzPopularEnsembleMethods1999, zhangEnsembleMachineLearning2012}, the results generated from various applied voltages can be combined to yield an ensemble prediction. This operation achieves a higher accuracy than that based on individual voltages, and outperforms the prediction based on even the optimal voltage alone.

Here we simply averaged the predicted probabilities from individual models with different applied voltages and made inferences based upon the highest average probability. The accuracy improvement for the three learning tasks is presented in Fig. 3b (Fashion MNIST test), Fig. 3c (MNIST handwritten digits test), and Fig. 3d (EMNIST letters test). The results from all three tasks show that optical ensemble learning enhances performance in classification. The improvement is particularly significant when the number of grains is low. With 4 grains optical ensemble learning significantly reduces classification error, and boosts the accuracy to $51.9\%$ (from $45.8\%$) for Fashion MNIST, $57.3\%$ (from $47.4\%$) for MNIST handwritten digits, and $33.9\%$ (from $26.8\%$) for EMNIST letters. Even at high grain numbers the enhanced performance from optical ensemble learning is noticeable, achieving an accuracy at 784 grains of $83.3\%$ (EMNIST letters), $96.4\%$ (MNIST handwritten digits), and $85.5\%$ (Fashion MNIST). The confusion matrix for the Fashion MNIST classification using 100 grains is shown in Fig. 3b. The ensemble learning accuracy is improved to $81.58\%$, while the highest accuracy under the optimal voltage is $79.4\%$ (4.7V). These results demonstrate the effectiveness of the optical ensemble learning enabled by reconfigurable structural nonlinearity, which harnesses the nonlinear optical processing at all applied voltages, not just a single optimal voltage, to improve the inference accuracy.

To achieve optical ensemble learning, the repeatability of the optical system is crucial: the same speckle pattern must be consistently reproduced at the same applied voltage given the same input. This repeatability ensures that each trained digital readout layer can reliably process the speckle patterns generated under the corresponding voltage as the applied voltage cycles through a series of predetermined values $V_1$,$V_2$,…,$V_n$. In addition, the stability and repeatability are also important for any practical application and reliable system implementation and deployment, for example, when there is a significant time lapse between the training and the testing phases. To evaluate the repeatability of the optical system, we performed six speckle-recording experiments (Fig. 4, A1 $\sim$ C2). In each experiment, 60,000 Fashion MNIST images were sequentially encoded on the SLM and the corresponding nonlinearly mapped speckle patterns were recorded (consisting of 576 grains). Each experiment lasted about six hours, during which the voltage amplitude across the LCPC was maintained at the same value. We cycled through applied voltage amplitudes on the LCPC from 2.6V, 4.7V to 7.6V during the first three experiments (referred to as $A_1$,$B_1$,and $C_1$ tests, respectively). To minimize any hysteresis effect, the applied voltage was first reset to zero before changing to a new value. After a few hours of intervening gap, the process was repeated to perform recordings at the same three voltages, yielding three additional tests $A_2$,$B_2$,and $C_2$. In other words, at each voltage we have two sets of speckle patterns recorded at about 20 hours apart (2.6 V: $A_1$ and $A_2$, 4.7V: $B_1$ and $B_2$, 7.6V: $C_1$ and $C_2$).

\begin{figure}[H]
    \begin{subfigure}{\textwidth}
        \centering
        \includegraphics[width=\textwidth]{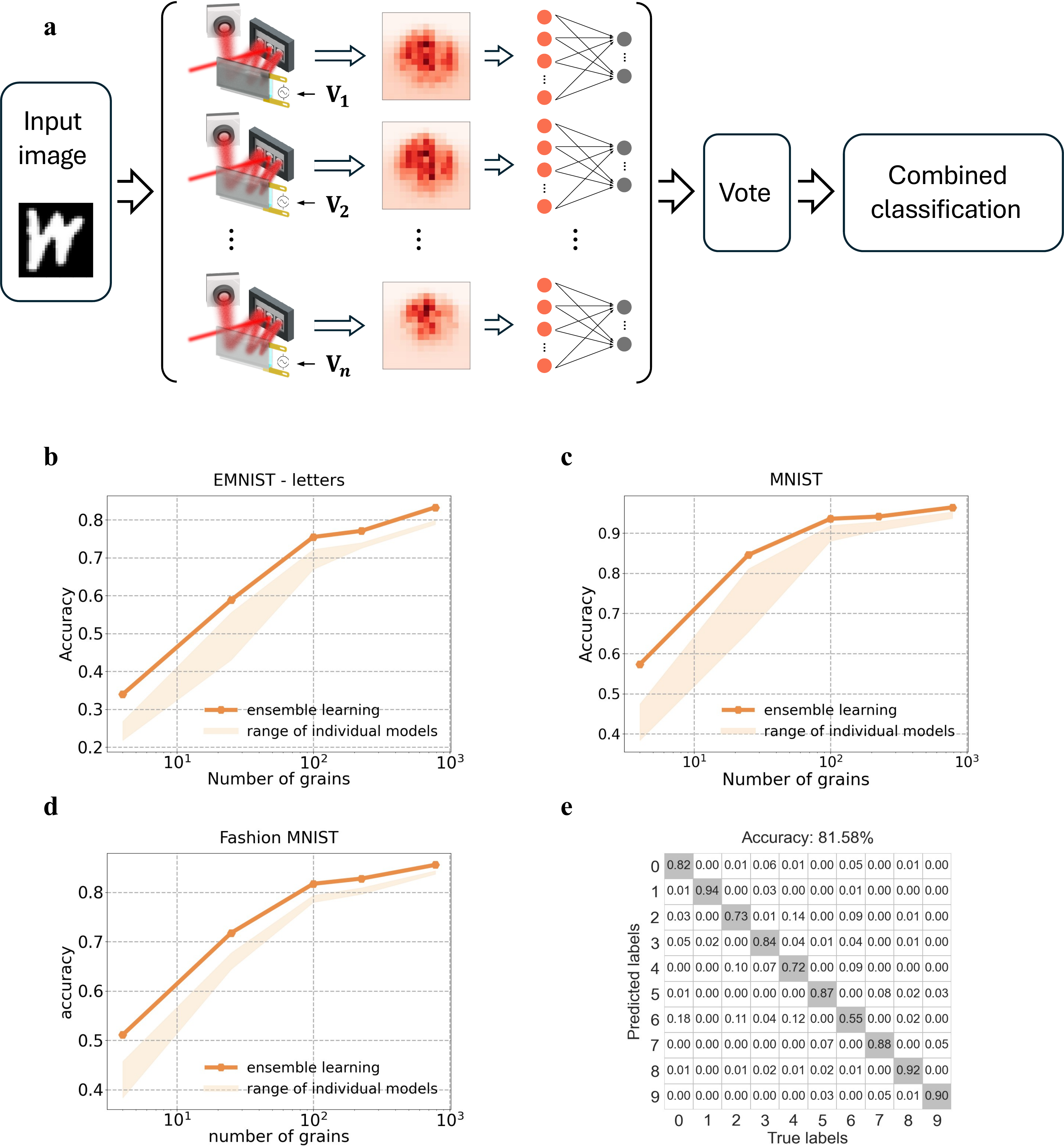}
    \end{subfigure}
    \label{fig:3}
\end{figure}
\clearpage
\begin{figure}[ht]\ContinuedFloat
  \centering
    \caption{\textbf{Optical ensemble learning. (a)} Schematic diagram of the reconfigurable structural nonlinearity based optical ensemble learning, which combines the results generated by various applied voltages to yield a higher prediction accuracy. The ensemble learning accuracies for \textbf{(b)} EMNIST letters, \textbf{(c)} MNIST hand-written digits and \textbf{(d)} Fashion MNIST classification concerning different numbers of grains (solid curve). For comparison, the range of accuracies obtained with individual voltages is also shown (shaded area). In all three cases, the reconfigurable structural nonlinearity based optical ensemble learning enhances the performance in classification. The improvement is particularly significant when the number of grains is low, for example in \textbf{(c)}, boosting the accuracy from $47.4\%$ to $57.3\%$ in MNIST handwritten digits classification with only 4 grains. \textbf{(e)} Confusion matrix for the Fashion MNIST classification using optical ensemble learning, which combines the results from the applied voltages of 2.6V, 4.7V, and 7.6V. 100 grains are used. The ensemble learning accuracy is improved to $81.58\%$, while the highest accuracy under the optimal voltage is $79.4\%$ (4.7V).}
\end{figure}

First, we compared the speckle patterns generated by the same input images, at the same applied voltage but from different experiments (i.e., $A_1$ vs. $A_2$, $B_1$ vs. $B_2$, $C_1$ vs. $C_2$). The structural similarity index measure (SSIM) was used to quantify each pair of patterns. As shown in the histograms (Fig. 4a, b, c), the speckle patterns produced under the same applied voltage show high repeatability, having an average SSIM above $98\%$ in all three cases. The smallest voltage (2.6 V) shows the best consistency among the three voltages studied here, producing the narrowest histogram distribution with a mode higher than 0.99. Next, we performed cross-validation testing using these pairwise datasets (Fig. 4d, e, f): the testing speckle dataset from one experiment was used to evaluate the readout layer trained using the training speckle dataset in the same or the other experiment (under the same applied voltage). On average, when the training and testing datasets come from different experiments, the model accuracy is less than $1\%$ lower than that when both come from the same experiment. These results indicate that our reconfigurable nonlinear optical processing system maintains high repeatability and robustness for optical ensemble learning.
\begin{figure}[H]
    \begin{subfigure}{\textwidth}
        \centering
        \includegraphics[width=\textwidth]{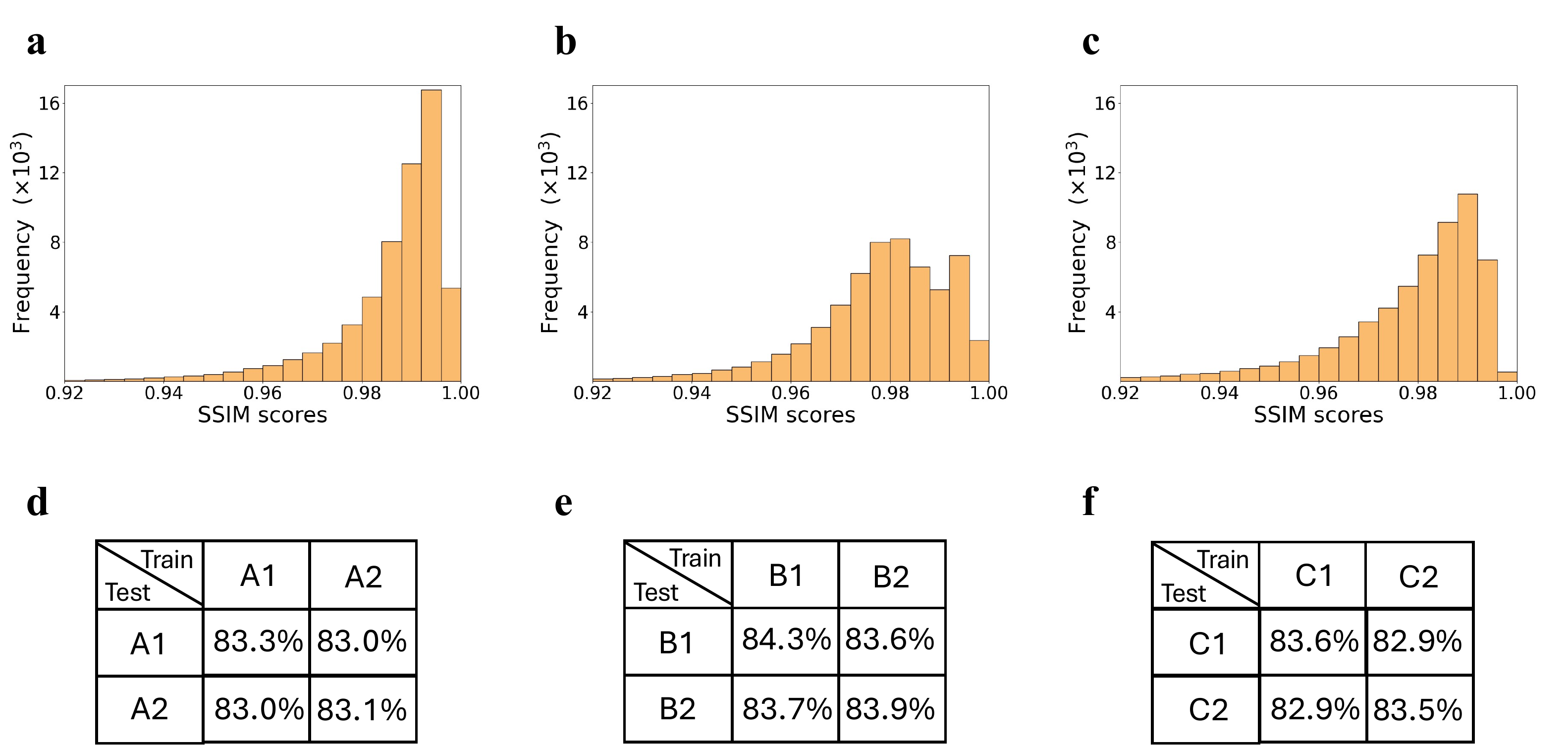}
    \end{subfigure}
    \label{fig:4}
    \caption{\textbf{Cross-validation results showing the repeatability of the reconfigurable nonlinear optical processing system.} Six speckle recording experiments were performed (denoted as $A_i$, $B_i$, and $C_i$, $i=1,2$). In each experiment, 60,000 Fashion MNIST images were sequentially encoded on the SLM, and the corresponding nonlinearly mapped speckle patterns were recorded (consisting of 576 grains). The voltage applied to the LCPC was kept at 2.6V, 4.7V and 7.6V in A1 and A2, B1 and B2, and C1 and C2, respectively. \textbf{(a) / (b) / (c)} Histograms showing SSIM values between pairwise speckle patterns generated by the same input images at the same applied voltages but from different experiments: A1 vs. A2 \textbf{(a)}, B1 vs. B2 \textbf{(b)}, and C1 vs. C2 \textbf{(c)}. \textbf{(d) / (e) / (f)} Cross-validation results between experiments done at the same applied voltages. Testing speckle patterns from one experiment were used to evaluate the accuracy of the readout layer trained using the training speckle patterns from the same or the other experiment.}
\end{figure}

\section{3. Discussion}
We have introduced an ensemble nonlinear optical learner. Both the nonlinear mapping between the input digital data and the resultant speckle patterns and its reconfigurability rely on multiple linear optical scatterings rather than material nonlinearity, leading to simplicity in optical implementation with low-power light sources. The total scattering potential of this system ($V$) can be partitioned into two terms, one of which represents the input information ($V_I$) implemented by the SLM, while the other arises from the tunable LCPC ($V_C$):
\begin{equation}
    V=V_I + V_C \label{eqn:1}
\end{equation}
The nonlinear mapping between the scattering potential and the output scattered field can be modeled using Born’s series\cite{eliezerTunableNonlinearOptical2023}: 
\begin{equation}
    \begin{aligned}
        \mathbf{E_{sc}}&= GV\mathbf{E_{in}} + {[GV]}^2\mathbf{E_{in}} + {[GV]}^3\mathbf{E_{in}} + ...\\
        &=G(V_I + V_C)\mathbf{E_{in}} + {[G(V_I + V_C)]}^2\mathbf{E_{in}} + {[G(V_I + V_C)]}^3\mathbf{E_{in}} + ...\\
        &=[GV_I + GV_C + GV_IGV_I + GV_IGV_C + GV_CGV_I + GV_CGV_C + GV_IGV_IGV_C + ...]\mathbf{E_{in}}\\
        &=T\mathbf{E_{in}}  \label{eqn:2}
    \end{aligned}
\end{equation}
In Equation~\ref{eqn:2}, $\mathbf{E_{in}}$ and $\mathbf{E_{sc}}$ are the input and the scattered fields, $G$ is the operator of free-space Green’s function, $V_C$ and $V_I$ are operators of the control potential of the LCPC and the information-bearing scattering potential encoded by the SLM, respectively, and T is the transfer operator discussed earlier. The expansion begins with the first order term $G(V_I+V_C) \mathbf{E_{in}}$ denoting the single-scattering contribution. Subsequent terms show high-order scatterings. Note that the relationship between $\mathbf{E_{sc}}$ and $\mathbf{E_{in}}$ is linear, meaning the system does not rely on any nonlinear optical effects and thus there is no need for high power pulsed lasers\cite{eliezerTunableNonlinearOptical2023, liNonlinearEncodingDiffractive2024}.

The nonlinear relationship between $V_I$ and $\mathbf{E_{sc}}$ can be clearly seen in Equation~\ref{eqn:2}, where high-order terms such as $GV_IGV_CGV_I...GV_CGV_I \mathbf{E_{in}}$ give rise to a nonlinear mapping between the input data encoded by $V_I$ and $\mathbf{E_{sc}}$. Importantly, this nonlinear mapping can be reconfigured by controlling $V_C$, providing a mechanism to realize an ensemble of nonlinear processors for enhancing the learning performance.

Our system was applied to three image classification tasks; by tuning the structural nonlinearity, it achieved an accuracy of $84.3\%$ in the Fashion MNIST test, $95.3\%$ in the MNIST hand-written digits test, and $79.9\%$ in the EMNIST letters test, respectively, with 900 speckle grains or super-pixels. Enhanced data compression capability was also demonstrated: using only 4 grains, we obtained accuracies of $45.8\%$ (Fashion MNIST), $47.4\%$ (MNIST hand-written digits), and $26.8\%$ (EMNIST letters) under their respective optimal voltages. Our studies showed that using optical ensemble learning, classification accuracies with only 4 grains can be significantly improved to $51.9\%$, $57.3\%$, and $33.9\%$ respectively - higher than the accuracies obtained under optimal applied voltages. Owing to these attributes enabled by reconfigurability in nonlinear mapping, our approach can be particularly advantageous in edge computing applications as it requires only limited electronic computational resources to train single-layer networks and few sensors/detectors (matching the speckle grain number). Therefore, our work can potentially help accelerate the development of next-generation intelligent optical sensors and imagers.

\section{4. Materials and Methods}

\subsection{4.1 LCPC sample preparation}
The liquid crystal/polymer composite (LCPC) mixture comprises $80.0wt\%$ E7, a nematic liquid crystal from HCCH, and $20.0wt\%$ NOA65 from Norland. It was enclosed within a glass cell made of two ITO glass substrates spaced $8\mu m$ apart. The substrates were bonded along two opposite edges using an adhesive with $8\mu m$ glass spheres serving as spacers, while the other two sides remained unsealed to allow for LCPC injection. Before the injection, the mixture was heated to approximately $90^{\circ} C$ to achieve its isotropic phase and was maintained at this temperature throughout the injection process. The mixture was injected into the glass cell using capillary action, and it transitioned back to the liquid crystal phase upon cooling down to the room temperature. Then, the LCPC cell was illuminated by $365nm$ unpolarized UV light (Brightek BK Spotcure 100UV) at the intensity of $15mW/cm^2$ for 30 minutes to induce phase separation (between liquid crystal and polymer) and photo-polymerization. After the illumination process was completed, the two remaining openings of the cell were sealed. For capturing polarized micrographs of prepared LCPC samples, a polarized optical microscope (Eclipse LV100 POL from Nikon) was utilized, coupled with a digital camera (DS-Fi1 from Nikon). The white light source used for the microscope is a halogen light bulb (64410 HLC from Osram).

\subsection{4.2 Optical experimental setup}
A multi-bounce configuration formed by a mirror-backed LCPC cell and a spatial light modulator (SLM) was used to realize structural nonlinearity. The scattering characteristics of the LCPC was controlled by a sinusoidal voltage signal at 1 kHz (generated by an Agilent 3320A function generator and amplified with a Trek model 2220 high voltage amplifier), which was applied across the two electrodes of the cell. In the characterization of reconfigurable structural nonlinearity using Boolean analysis, a digital mirror device (DMD, Texas Instruments DLP6500FYE) was used due to the requirement of binary input data. In the image classification tasks, a liquid crystal based SLM (Holoeye Photonics PLUTO-2.1, resolution 1920×1080, pixel pitch $8.0\mu m$) was used to encode input images. During the experiments, an incident He-Ne laser beam (Laser Research model \#5mW, wavelength 633nm) bounced between the LCPC and the SLM, which were positioned at an angle (see Fig. S3 in supplementary materials). After multiple scatterings by the LCPC and the SLM, the output speckle pattern was recorded by a camera (Lumenera Lu100M, $1280\times1024$ resolution).

\subsection{4.3 Characterization of reconfigurable structural nonlinearity}
The DMD was used to encode $3\times3$ binary input data. The active area of the DMD has $1920\times1080$ micromirror pixels with a pixel pitch of  $7.56\mu m$. Each micromirror pixel can be tilted by $\pm12^{\circ}$, representing binary “on” or “off” states. With a $3\times3$ data format there are 512 (i.e., $2^{(3\times3)}$) different binary images in total. Each input image is up sampled to 300×300 and repeatedly displayed 3 times on the DMD. The nonlinear relationship between the output speckle patterns and the input binary images are expanded using the Boolean analysis, also known as the Fourier expansion.

\subsection{4.4 Image classification tests}
In the experiments evaluating the learning performance of the system with EMNIST letters, MNIST handwritten digits, and Fashion MNIST image classification tasks, each input was displayed three times at three different locations on the SLM as $280\times280$ phase images. The grayscale intensities (0-255) of the original image are linearly and proportionally mapped to phase values between 0 and $\pi$. Because the original dimension of these images is $28\times28$, each pixel in the original image was mapped to $10\times10$ SLM pixels, in order to match the image size with the incident beam size. In Fashion MNIST classification experiments, amplitudes of the voltages applied to the LCPC cell were 2.6V, 4.7V and 7.6V, while in MNIST and EMNIST letters classification experiments these were 2.6V, 4.7V, 7.6V, 10.0V and 12.0V. In each experiment, a sinusoidal voltage signal with a fixed amplitude was applied, and 60,000 input images (50,000 training set images and 10,000 test set images) were sequentially displayed on the SLM, with their corresponding speckle patterns recorded by the camera. The speckle patterns were then down-sampled; each “super pixel” (also called a grain) of the down-sampled image is the average of 32×32 camera pixels. This choice of super pixel size was determined by the mean speckle size of all patterns captured by the camera, estimated from the autocorrelation peak widths. The number of grains was controlled by selecting a cropping window of different sizes in the down-sampled speckle patterns, forming feature vectors with different lengths (i.e., 4, 25, 100, 225, 576, and 784). A single-layer fully connected linear network that directly linked the input neurons (i.e., values in feature vectors) to classification categories was trained using the 50,000 feature vectors that correspond to the training dataset images. All the single-layer linear networks were trained using the TensorFlow framework (version 2.6.0). The optimizer employed was ADAM, and the loss function used was sparse categorical cross entropy.

\subsection{Disclaimer}
Any options, finding, and conclusions or recommendations expressed in this material are those of the author(s) and do not necessarily reflect the views of the United States Air Force.
\subsection{Acknowledgments}
This material is based upon work supported by the Air Gorce Office of Scientific Research under award number FA2386-24-1-4054 (Z. Liu, X. Ni, and I. C. Khoo). The authors also acknowledge the support of the National Science and Technology Council of Taiwan NSTC113-2112-M-110-020-MY3, and NSTC113-2124-M-110-002-MY3 (T. H. Lin).
\subsection{Data availability}
The data and code supporting the plots within this paper and other findings of this study are available from the corresponding authors upon reasonable request.

% \begin{acknowledgement}

% Please use ``The authors thank \ldots'' rather than ``The
% authors would like to thank \ldots''.

% The author thanks Mats Dahlgren for version one of \textsf{achemso},
% and Donald Arseneau for the code taken from \textsf{cite} to move
% citations after punctuation. Many users have provided feedback on the
% class, which is reflected in all of the different demonstrations
% shown in this document.

% \end{acknowledgement}


\bibliography{snl}

\clearpage
\section{Supplementary Information}
\subsection{Liquid crystal/polymer composite (LCPC) as a tunable scattering layer}
LCPC mixture is enclosed by two ITO-coated glass substrates separated by an $8\mu m$ gap; the mixture constituents are $80wt\%$ E7 (a nematic liquid crystal from HCCH) and $20wt\%$ NOA65 (from Norland). Owing to the refractive index difference between the birefringent liquid crystals and the polymer, multiple interfaces with mismatched refractive indices are created within the heterogeneous morphology, leading to optical scattering. By electrically changing the orientation of liquid crystal molecules, these LC/polymer interfaces and the interface refractive index difference can be tuned by an applied voltage on the order of a few volts, affecting not only the intensity but also the wavefront and phase distribution of light passing through the LCPC.
\setcounter{figure}{0}
\renewcommand{\thefigure}{S\arabic{figure}}
\begin{figure}[H]
    \begin{subfigure}{\textwidth}
        \centering
        \includegraphics[width=0.86\textwidth]{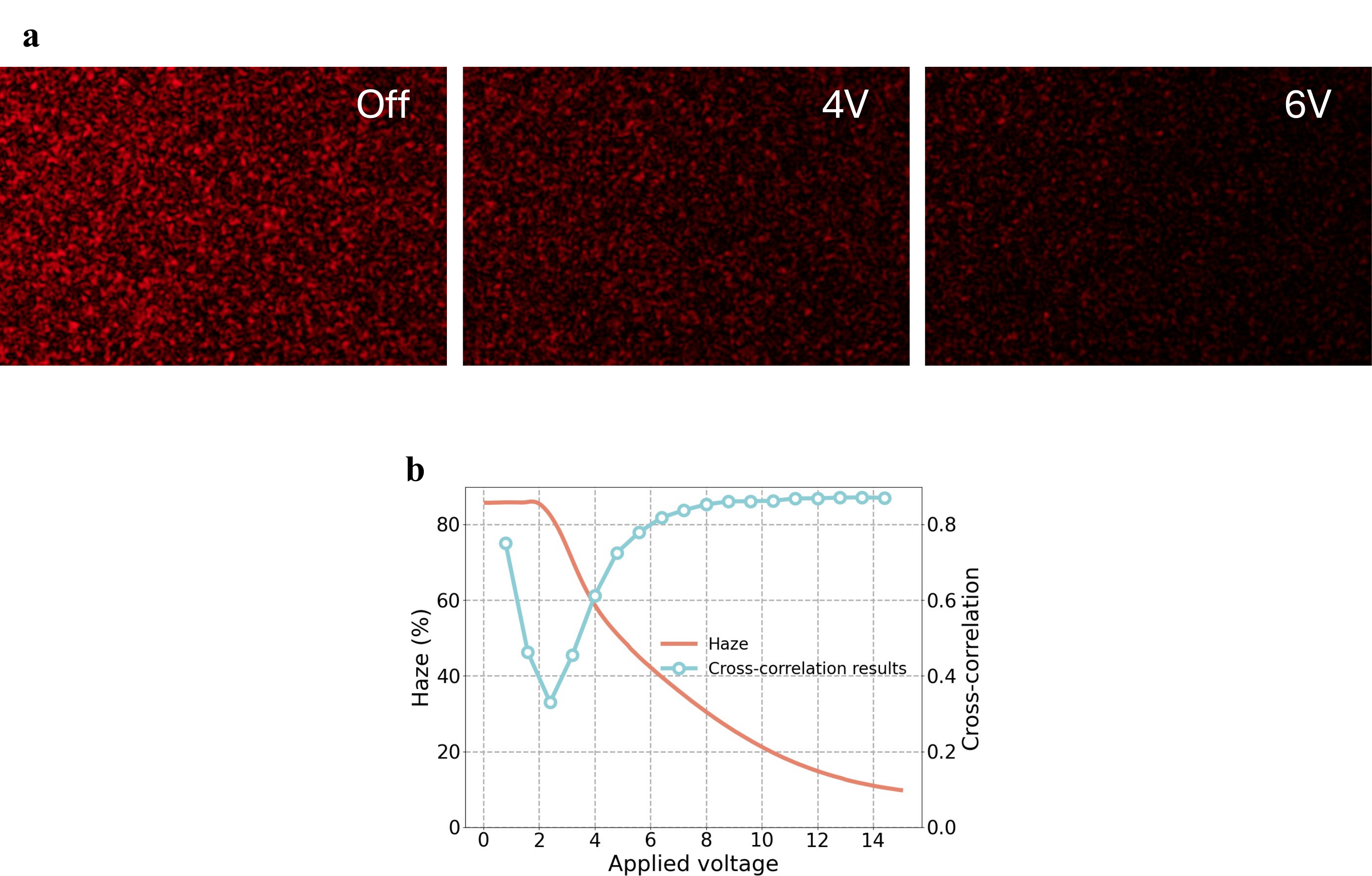}
    \end{subfigure}
    \label{fig:s1}
    \caption{\textbf{(a)} Scattering patterns of a He-Ne laser beam ($\lambda=633nm$) produced by the LCPC under different voltages. \textbf{(b)} Haze and the cross-correlation values between adjacent scattering patterns at voltage increments of 0.8V.}
\end{figure}
Fig. S1a shows changes in the scattering patterns of a 633nm He-Ne laser beam passing through the LCPC under different voltages. The LCPC sample was placed between a He-Ne laser (N-STP-912 from Newport) and a digital camera (E3ISPM from TOUPCAM). The distance between the sample and the camera was approximately 11.5cm. The speckle patterns were captured approximately 2cm away from the original beam axis. Fig. S1 clearly demonstrates that scattering characteristics such as haze and cross-correlations between adjacent speckle patterns can be electrically tuned. Haze is the ratio between the beam power (that emerges in directions not along the incident beam’s) and the total transmitted beam power; it is a measure of the opaqueness of the LCPC due to scattering. Here the haze of the LCPC cell is quantified using a haze meter (Nippon Denshoku Industries NDH7000). The cross-correlation values between adjacent speckle patterns at voltage increments of 0.8V quantify the similarity between the corresponding scattering patterns and therefore represent the change of scattering property by voltage tuning. As shown in Fig. S1b, in the range of 0-2V, the cross-correlation value drops dramatically as the haze remains constant, indicating a significant change in the scattering property while keeping a high opaqueness simultaneously. However, at higher voltages, the cross-correlation value increases with the voltage while the haze gradually decreases, implying that as the LCPC becomes more transparent, changing the voltage provides smaller modulations in the scattering property.

It is noteworthy that the field-driven scattering modulation of the LCPC exhibits reproducibility and stability. We repeatedly (30 times) apply 2V to the LCPC, with each application lasting 5s ($\tau_{on}=5s$), followed by a 5s interval ($\tau_{off}=5s$). We calculate the cross-correlation between the speckle pattern obtained during each voltage application and the initial speckle pattern (Speckle \#1) from the first voltage application. As shown in Fig. S2a, during the repeated voltage application process, the cross-correlation values remain at about 0.98, so every speckle pattern in this queue is similar to the initial speckle pattern, indicating high reproducibility. After 50 minutes of idling time, the same procedure is repeated, and its corresponding cross-correlation curve also remains around 0.98, showing good stability (Fig. S2b).
\begin{figure}[H]
    \begin{subfigure}{\textwidth}
        \centering
        \includegraphics[width=0.8\textwidth]{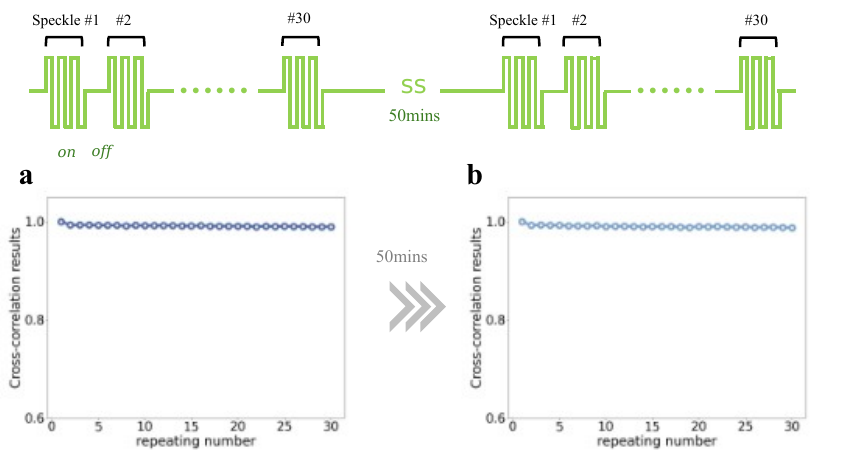}
    \end{subfigure}
    \label{fig:s2}
    \caption{\textbf{(a)} Cross-correlations between the speckle pattern obtained at each voltage application at $2V_{p-p}$ in a repeated voltage-application process and the initial pattern. \textbf{(b)} The same procedure was repeated after 50 minutes.}
\end{figure}

\begin{figure}[H]
    \begin{subfigure}{\textwidth}
        \centering
        \includegraphics[width=0.6\textwidth]{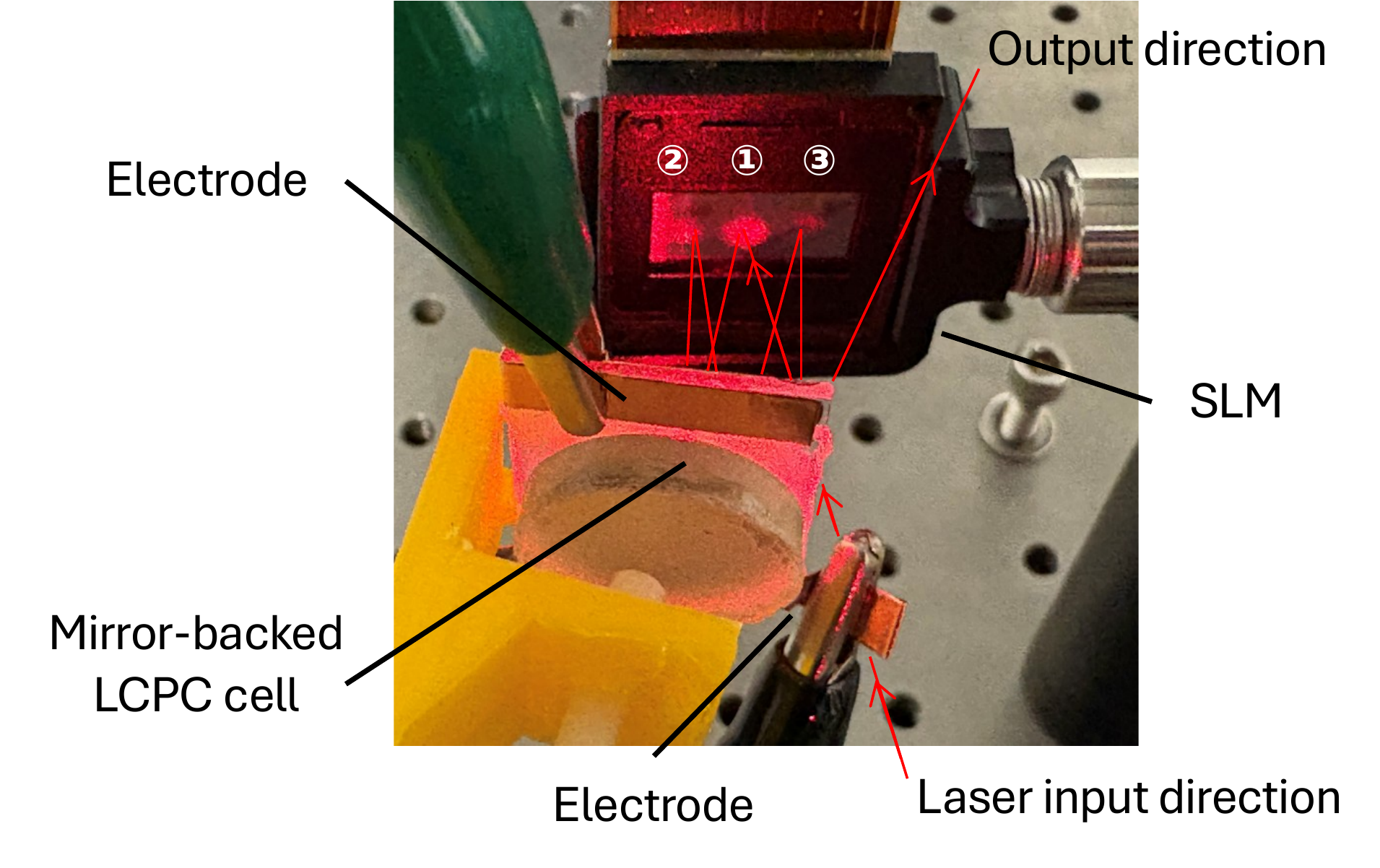}
    \end{subfigure}
    \label{fig:s3}
    \caption{\textbf{Photo showing the arrangement of the LCPC cell and the SLM.} A silver mirror was fixed at the backside of the LCPC cell by a 3D printed mount. The SLM was positioned opposite to the LCPC cell. A low-power, continuous-wave He-Ne laser was incident on the SLM, and subsequently underwent multiple scatterings by the SLM and the LCPC (as indicated by the schematic rays showing the beam propagation trajectory and three reflection points on the SLM screen). The output speckle pattern was recorded by a camera.}
\end{figure}

\subsection{Optical experimental setup used for image classification tests}
Fig. S3 is a photo of the experimental set up, showing the mirror-backed LCPC cell and the spatial light modulator (SLM, Holoeye Photonics PLUTO-2.1, resolution $1920\times1080$, pixel pitch $8.0\mu m$). A sinusoidal voltage signal at $1 kHz$ (generated by an Agilent 3320A function generator and amplified with a Trek model 2220 high voltage amplifier) is applied across the two electrodes of the LCPC cell (copper strips in Fig. S3). A continuous wave, low power He-Ne laser (Laser Research model \#5mW, wavelength 633nm) is incident on the SLM and scattered multiple times by the LCPC cell and the SLM. Schematic rays are provided to illustrate the beam bouncing trajectory. The output speckle pattern is recorded by a camera (Lumenera Lu100M, $1280\times1024$ resolution, not shown).

\subsection{Learning performance in image classification tasks}
The learning performance of the optical system is evaluated using three image classification benchmarks: EMNIST letters, MNIST hand-written digits, and Fashion MNIST. For each benchmark, 50,000 training samples and 10,000 test samples are utilized. Voltages of 2.6V, 4.7V, and 7.6V are applied to the LCPC in Fashion MNIST tests, while higher voltages (10.0V and 12.0V) are additionally applied to the LCPC in MNIST and EMNIST experiments. The captured speckle patterns are down sampled and cropped to form feature vectors. A single-layer linear network consisting of an input layer (number of input neurons equal to number of features) and an output layer (number of output neurons equal to number of classification categories, i.e., 10 classes for Fashion MNIST and MNIST hand-written digits, 26 classes for EMNIST letters) is trained using an ADMM optimizer to classify the input speckles.

The down-sampling is guided by the average size of speckle grains. The number of grains (also referred as super-pixels) in feature vectors varies from 784 to 4. For tests with 100 grains, the learning curves of accuracy and loss function (sparse categorical cross entropy) for Fashion MNIST, MNIST hand-written digits, and EMNIST letters are shown in Fig. S4a, S4b, and S4c, respectively. These results demonstrate that the voltage applied to the LCPC can impact the classification accuracy for the same classification task. For example, each task has an optimal voltage for the case of 100 grains: 4.7V for Fashion MNIST ($79.40\%$ accuracy), 10.0V for MNIST hand-written digits ($91.92\%$) and EMNIST letters ($72.28\%$). Confusion matrices for MNIST hand-written digits and EMNIST letters tests (100 grains, 10.0V) are presented in Fig S4d and S4e. 

Additional examples of the analysis of the Euclidean distances between feature vectors in EMNIST letters experiments is shown in Fig. S4f and S4g. Fig. S4f shows the distinction between intra-class (between speckle patterns within the class ‘i’) and inter-class distance distributions (between speckle patterns in class ‘i’ and class ‘o’) is greater at 2.6V (K-L divergence 6.31) than at 12V (K-L divergence 5.05). On the other hand, Fig. S4g reveals an opposite trend between class ‘c’ and class ‘x’, where 12V (K-L divergence 4.50) outperforms 2.6V (K-L divergence 2.29) in terms of class separation.

The corresponding confusion matrices using photonic ensemble learning are shown in Fig. S4h and S4i, where the classification accuracy for MNIST hand-written digits and EMNIST letters are improved to $93.58\%$ and $75.41\%$, respectively. These results underscore that the reconfigurability enabled by the LCPC can be leveraged to enhance the learning performance.
\clearpage
\begin{figure}[H]
    \begin{subfigure}{\textwidth}
        \centering
        \includegraphics[width=\textwidth]{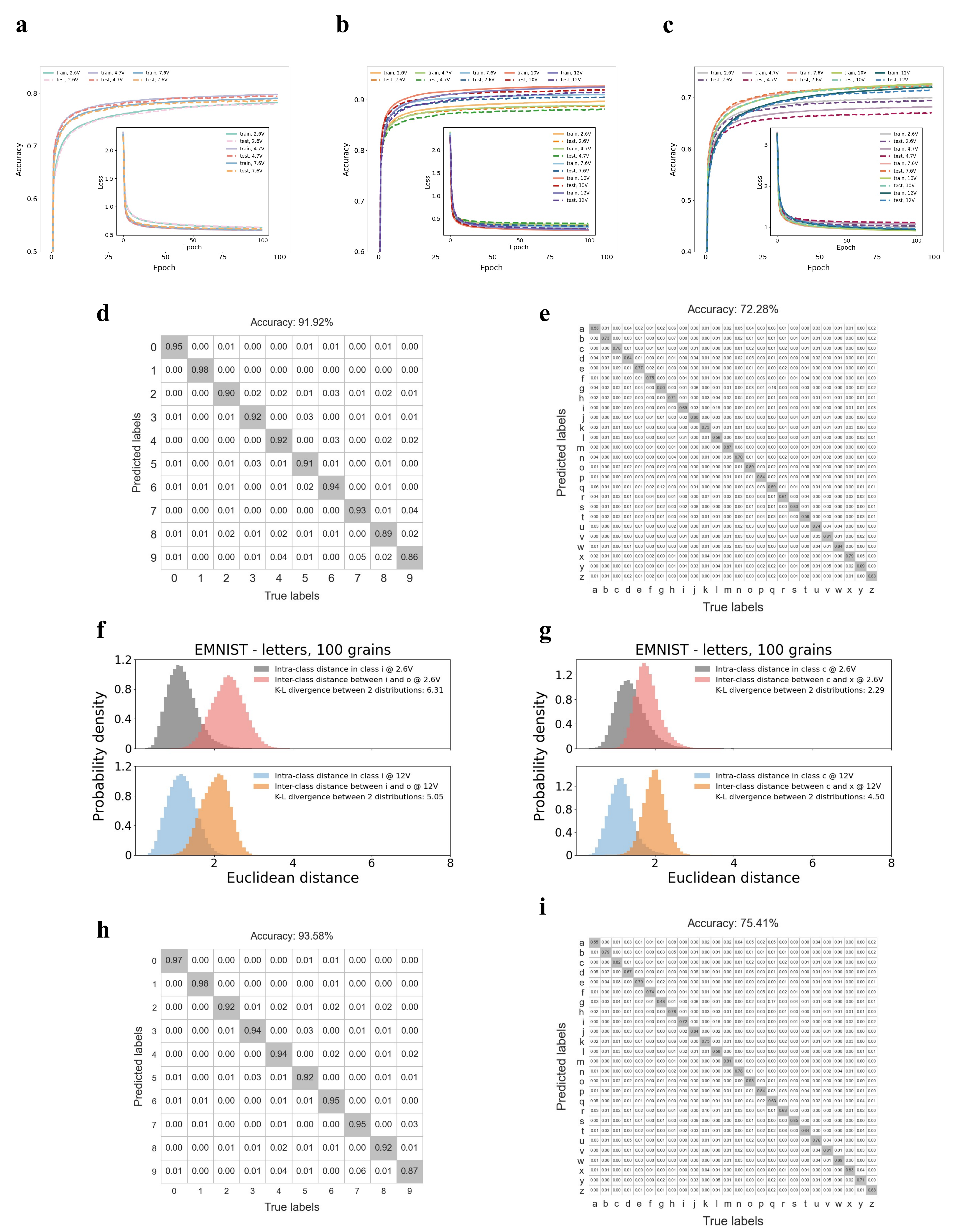}
    \end{subfigure}
    \label{fig:s4}
\end{figure}
\clearpage
\begin{figure}[ht]\ContinuedFloat
  \centering
    \caption{\textbf{Learning curves of accuracy and loss function for: (a)} Fashion MNIST dataset classification under 3 different voltages, \textbf{(b)} MNIST dataset classification under 5 different voltages, and \textbf{(c)} EMNIST letters dataset classification under 5 different voltages. In the legends, “train” refers to (accuracy or loss for) training datasets and “test” refers to test datasets; voltage values are amplitudes of alternating-current control signals applied to the LCPC. \textbf{(d)} Confusion matrix for MNIST hand-written digits classification with 100 speckle grains and 10.0V voltage amplitude applied to the LCPC. \textbf{(e)} Confusion matrix for EMNIST letters classification with 100 speckle grains and 10.0V voltage. \textbf{(f)} Histograms showing the distributions of intra-class Euclidean distances between feature vectors (with 100 grains) within EMNIST letters dataset class ‘i’ and inter-class Euclidean distances between feature vectors in class ‘i’ and class ‘o’, under 2.6V and 12V. \textbf{(g)} Histograms showing the distributions of intra-class Euclidean distances between feature vectors within EMNIST letters dataset class ‘c’ and inter-class Euclidean distances between feature vectors in class ‘c’ and class ‘x’, under 2.6V and 12V. The distinction between intra- and inter-class distances distributions under the same voltage is quantified by Kullback-Leibler divergence. \textbf{(h)} Confusion matrix for MNIST handwritten digits classification (100 grains) using photonic ensemble learning. \textbf{(i)} Confusion matrix for EMNIST letters classification (100 grains) using photonic ensemble learning.}
\end{figure}

\end{document}